# A Novel Algorithm for String Matching with Mismatches

**Vinodprasad P.**
*Sur College of Applied Sciences, Ministry of Higher Education, Sultanate of Oman*
*vinod.sur@cas.edu.om*

Abstract: We present an online algorithm to deal with pattern matching in strings. The problem we investigate is commonly known as 'string matching with mismatches' in which the objective is to report the number of characters that match when a pattern is aligned with every location in the text. The novel method we propose is based on the frequencies of individual characters in the pattern and the text. Given a pattern of length M, and the text of length N, both defined over an alphabet of size σ, the algorithm consumes *O(M)* space and executes in *O(MN/σ)* time on the average. The average execution time *O(MN/σ)* simplifies to *O(N)* for patterns of size M ≤ σ. The algorithm makes use of simple arrays, which reduces the cost overhead to maintain the complex data structures such as suffix trees or automaton.

# 1 INTRODUCTION

Similarity search is a fundamental problem in pattern recognition. Similarity searches allow for some mismatches between the text and the pattern. Searching for similar patterns is common in DNA sequence analysis, data mining, search engines, and many other applications. The term 'distance' is used quite often when comparing two strings for similarity. One of the simplest distance-metric is the Hamming distance. The hamming distance between two equal length strings is the number of mismatch symbols at corresponding locations. In literature, this problem is sometimes also called '*string matching with k-mismatches*'.

## 1.1 String Matching with K Mismatches

Let R and S be two non-empty equal-length strings of size M such that $R = r_0\ r_1\ ...r_{M-1}$ and $S = s_0\ s_1....s_{M-1}$. Then, the Hamming distance between R and S is given by *ham(R, S)* = number of all locations $i$ where $r_i \neq s_i$ such that $0 \leq i \leq M-1$. Modern applications require large databases to be searched for regions that are similar to a given pattern. In such a context, the '*k-mismatch'* problem can be stated as follows:

*Given the text* $T = t_0\ t_1...t_{N-1}$ *of size N, and the pattern* $P = p_0\ p_1....p_{M-1}$ *of size M such that* $M \leq N$. *Both text and the pattern are defined over alphabet λ. Let* $hd_i$ *be the Hamming distance such that* $hd_i = ham(P,\ t_i\ t_{i+1}...t_{i+M-1})$, *where,* $0 \leq i \leq (N-M)$. *Then, for a given integer k such that* $0 \leq k \leq M$, *report all locations i in T where* $hd_i \leq k$.

To solve the '*k-mismatch'* problem, Landau and Vishkin (1986) proposed a suffix-tree-based algorithm. A suffix tree is created using the text and the pattern in *O(N+M)* time and space, before applying searches to report *k-mismatches* in *O(kN)* time. For suffix trees, see (McCreight, 1976), (Ukkonen, 1995), and (Gusfield, 1999). The Tarhio and Ukkonen (1993) algorithm requires the pattern to be preprocessed in *O(kσ)* space and *O(M+kσ)* time, and then reporting *k-mismatches* in *O(kN(k/σ +1/(M-k)))* time. The algorithm (Tarhio et al., 1993) is based on (Boyer and Moore, 1977), and (Horspool, 1980). The Galil and Giancarlo (1986) algorithm runs in *O(kN)* time and *O(M)* space. Amir, Lewenstein, and Porat (2004) provide $O(N\sqrt{k}\log k)$ time algorithm to solve the same problem.

## 1.2 String Matching with Mismatches

For k=M, the '*k-mismatch'* problem becomes independent of *k*, and hence can be stated as:

Given *the text* $T = t_0\ t_1....t_{N-1}$, *and the pattern* $P = p_0\ p_1....p_{M-1}$. *For every i in T such that* $0 \leq i \leq (N-M)$, *output the Hamming distance* $hd_i$ *such that* $hd_i = ham(P,\ t_i\ t_{i+1}...t_{i+M-1})$. In this case, the objective is to report all mismatches (0 to M) from every alignment location in the text. Therefore, the problem is commonly known as '*string matching* *with mismatches'*. Using a linked list, Baeza-yates and Perleberg (1996) proposed $O(N+Nf_{max})$ time and $O(2M + \sigma)$ space algorithm, where $f_{max}$ is the frequency of the most commonly occurring character in the pattern. Based on the Boolean convolution of the pattern and the text, Abrahamson (1987) solves the problem in $O(N\sqrt{(M \log M)})$ time and $O(N)$ space. Recently, Nicolae and Rajasekaran (2013) have shown that for pattern matching with wild-cards, the algorithm (Abrahamson, 1987) can be modified to obtain an $O(N\sqrt{(g \log M)})$ time, where $g$ is the number of non-wild-card positions in the pattern. The algorithm Clifford and Clifford (2007) is also based on convolution, which takes $O(N\log(M))$ time. A randomized algorithm Kalai (2002), which is based on (Karp and Rabin, 1987), consumes $O(N\log(M))$ time. Atallah, Chyzak, and Dumas (2001) approximate the number of mismatches from every alignment in $O(rN \log(M))$ time, where $r$ is the number of iteration algorithm has to make.

In literature, string-processing algorithms have made extensive use of suffix trees, suffix arrays, and automata. Most of these methods are covered Crochemore, Hancart, and Lecroq (2007). Algorithms based on automata give the best worst-case time $O(N)$. However, exponential time and space dependence on M and $k$ limits its practicality see Navarro (2001). Therefore, automaton based algorithms best suited to short patterns with low error rates Navarro (2001). Suffix trees on the other hand, consume space linear to the size of the text, which may be a challenge when dealing with the large text.

Breaking the trend, in this paper, we follow a novel approach to solve the 'string matching with mismatches' problem. The method we propose is based on the frequencies of individual characters in the pattern and the text, which is completely different from the other methods proposed in the past. The algorithm we propose avoids all complex data-structures, yet achieves average case $O(N)$ time for patterns of length $\leq \sigma$. The rest of the paper is structured as follows: - In section 2, we introduce few terms and notations used in this paper. To create a theoretical base for the algorithm, the lemma, corollaries, and examples are given in section 3. In section 4, an algorithm is provided to pre-process the pattern, which is a prerequisite for the main algorithm given in section 5. For a better understanding of the algorithm, the run-time behavior of the algorithm is also described in the same section. In section 6, we discuss the time and space requirements of the algorithms. Using real-life data, experimental results are provided in section 7. Finally, we conclude our work in section 8.

## 2 PRELIMINARIES

The symbol 'λ' represents the alphabet -a finite non-empty ordered set of characters, such that $|\lambda|=\sigma$ is the size of the alphabet. We use the symbols T and P to represent non-empty text and pattern strings of length N and M respectively. Both T and P are defined over the alphabet λ. T[i] or $t_i$ represents the $i^{th}$ character of T, where '$i$' is referred to as shift, location, or index in T. Throughout the paper, we have used a phrase extensively "*Number of matches of P at shift t in T"*, which refers to the total number of the characters that match when pattern P is aligned with shift $t$ in T. In the algorithm, we refer to this as the number of hits at shift $t$ in T by the pattern P. Note, for a clear relationship among the lemma, corollaries, examples, and the algorithms we consider the number of character matches (not mismatches).

## 3 LEMMA

Consider the text $T = t_0\, t_1\, t_2\, t_3\, t_4\, t_5 =$ DBCDAB of size N=6, and the pattern P = DABCD of size M=5. It is easy to see that one character match may be found provided that P is aligned at location -4 in T (assume that there is such a location). Similarly, a three character match may be found when P is aligned at locations -1 and 3 in T. Traditionally, pattern P is aligned with all locations $i$ in T such that $0 \leq i \leq N\text{-}M$. However, considering $i$*'s* in the extended range $(1\text{-}M) \leq i \leq (N\text{-}1)$ may also provide useful information, particularly when the pattern and the text are almost same in length, and the character matches exist at opposite ends of the strings being matched. Therefore, with the extended search space, the '*string matching with mismatches*' problem can be re-formulated as:-

*Given a text T and a pattern P. For every i in T such that $(1\text{-}M) \leq i \leq (N-1)$, output the Hamming distance $hd_i$ such that $hd_i = ham(P,\ t_i\, t_{i+1} \ldots t_{i+M-1})$, where, $t_i$ =null if $i < 0$ or $i > N\text{-}1$.* Now, for the text and the pattern given above, we are in a position to say that the hamming distance between P and $t_{-1}\, t_0\, t_1\, t_2\, t_3$ = 3, i.e., $hd_{-1} = ham(P,\ t_{-1}\, t_0\, t_1\, t_2\, t_3) = 3$. Similarly, $hd_3 = ham(P,\ t_3\, t_4\, t_5\, t_6\, t_7) = 3$. The algorithm given in section 5 solves the problem outlined above with the extended search space.

## 3.1 The Set Intersection Lemma

Let T be a text of length N, and P be a pattern of length M such that: T = T[0….N-1], and P = P[0….M-1]. For each shift $j$ in P, we define a set $R_j$ such that: $R_j = \{ i - j \mid T[i]=P[j], \forall 0 \le i \le N-1 \}$. Further, let S be a set such that $S = R_0 \cap R_1 \cap R_2 ... \cap R_{M-1}$. Then, every element $t \in S$ represents an exact match of P at shift $t$ in T, and the cardinality |S| represents the number of occurrences of P in T.

**Proof:** The given lemma has a simple and straightforward proof. Let P is present in T at shift $t$. Then, we have to show that $t \in S$. Let P appears in T at shift $t$, that means all M characters of pattern P = P[0…M-1] can be successfully matched with T[t, t+1…t+M-1]. Hence, $\forall j$ in P, T[ t+j] = P[j]. Now, from the definition of $R_j$, $\forall j$ in P we have: $R_j = \{ ( t + j ) - j \} = \{ t \} \Rightarrow$ for all $j$ in P, we have $t \in R_j \Rightarrow t \in S$. Further, since $t \in S$ represents an exact match of P in T at shift t $\Rightarrow$ |S|= Number of occurrences of P in T. Therefore, if $R_0 \cap R_1 \cap R_2 ...... \cap R_{M-1} = S = \{ \}$ then, exact match of P is not available in T. Notice, since $0 \le i \le N-1$, and $0 \le j \le M-1 \Rightarrow$ each element of $R_j$ lie in the range $(1-M) \le t \le N-1$.

**Example 1:** Let T = CABABABCBA be a text array of size 10, and P = ABAB be a pattern array of size 4.

For each shift $j$ in P, we create a set $R_j$ such that $R_j = \{ i - j \mid T[i] = P[j], \forall 0 \le i \le 9 \}$. Which gives: $R_0$ = {1, 3, 5, 9}, $R_1$ = {1, 3, 5, 7}, $R_2$ = {-1, 1, 3, 7}, and $R_3$ = {-1, 1, 3, 5}. Hence, $R_0 \cap R_1 \cap R_2 \cap R_3 = S$ = {1, 3}. Which confirms two (|S|) occurrences of P in T at shift 1, and 3.

**Corollary 1:** Given the M sets $R_j$ defined as above. Let $f_t$ be the frequency of occurrence of an integer '$t$' in all sets. Then, $f_t$ represents the number of characters that match at corresponding locations when P is aligned with shift $t$ in T.

**Proof:** As we proved already, $t \in S$ represents exactly M character matches of P at shift $t$ in T. therefore, each $t \in R_j$ represents a single character match of P $\Rightarrow$ the frequency of integer $t = f_t$ = Number of characters, that match at corresponding locations when P is aligned at shift $t$ in T.

**Example 2:** Consider example 1, the integer 5 appears in three sets: $R_0$, $R_1$, and $R_3$. Hence, the frequency of integer 5 = $f_5$ = 3. This confirms a three characters match of P when P is aligned with shift 5 in T $\Rightarrow$ hamming distance $hd_5 = M - f_5 = 4 - 3 = 1$. Similarly, $f_7$=2 and $f_{-1}$=2 reveal two characters match, when P is aligned with locations 7 and -1 in T respectively.

**Observation 1(a)**: As we proved in the lemma, for an exact match of P at shift $t$ in T, the integer $t$ must be present in all M sets, i.e., $f_t$ = M.

**(b)** Since, $f_t$ represents the number of characters that match $\Rightarrow M - f_t$ represents the Hamming distance, i.e. the number of mismatches, when the pattern P is aligned with shift $t$ in T i.e. $hd_t = ham$(P, T[t] T[t+1]...T[t+M-1] ) = $M - f_t$.

**Observation 2:** The lemma and examples given above may suggest that the method under consideration requires the entire text to be available before we create the sets. However, this is not the case. Consider example 1 again, to identify the first match at shift 1 in T, we do not require all elements of $R_j$ at once. Let us assume that we receive the text in the form of a stream of characters, and that we have just received T[0..5]. Applying the set intersection lemma for the given fragment of text, we get: $R_0$={1, 3, 5}, $R_1$={1, 3}, $R_2$={-1, 1, 3}, and $R_3$={-1,1}. Thus, we find that: $R_0 \cap R_1 \cap R_2 \cap R_3 = S$ = {1}, which confirms the first exact match of P=ABAB at shift 1 in T. Moreover, the double repetition of -1 and 3 reveals two characters match, when P is aligned at shifts -1 and 3 in T[0..5].

**Example 3:** Let's now discuss the practical aspect of the method. Consider the text T=SKRFCTHZCTZCFTYCTZGHTTCTHZTHZFC THZCTZC of size 38, and the pattern P = FCTHZCTZCF of size 10.

Table 1: Sets showing integer frequencies.

| Pattern Char | Shift in T *(i)* | Shift in P *(j)* | Set $R_j$ |
|---|---|---|---|
| F | 3, 12, 29, 38 | 0 | $R_0$ = **3**, 12, **29**, 38 |
| | | 9 | $R_9$ = -6, **3**, 20, **29** |
| C | 4, 8, 11, 15, 22, 30, 34, 37 | 1 | $R_1$ = **3**, 7, 10, 14, 21, **29**, 33, 36 |
| | | 5 | $R_5$ = -1, **3**, 6, 10, 17, 25, **29**, 32 |
| | | 8 | $R_8$ = -4, 0, **3**, 7, 14, 22, 26, **29** |
| T | 5, 9, 13, 16, 20, 21, 23, 26, 31, 35 | 2 | $R_2$ = **3**, 7, 11, 14, 18, 19, 21, 24, **29**, 33 |
| | | 6 | $R_6$ = -1, **3**, 7, 10, 14, 15, 17, 20, 25, **29** |
| H | 6, 19, 24, 27, 32 | 3 | $R_3$ = **3**, 16, 21, 24, **29** |
| Z | 7, 10, 17, 25, 28, 33, 36 | 4 | $R_4$ = **3**, 6, 13, 21, 24, **29**, 32 |
| | | 7 | $R_7$ = 0, **3**, 10, 18, 21, 26, **29** |

The first column of table 1 summarizes all unique pattern characters. The second and the third columns of the table represent the shifts of the corresponding pattern character in the text and in the pattern respectively. Each row of the last column represents set $R_j$ as defined in the lemma. As noted above, the frequency of occurrence $f_t$ of an integer '*t*' represents the number of characters that match when P is aligned at shift *t* in T. For example, $f_3= f_{29}$ = 10=M represent 10 characters match (exact match) of P at alignment locations 3 and 29 in T. Therefore, we simply need a mechanism to count the number of occurrences of individual *t's* in the last column of the table. This can be done using an array of size N, with all array cells having an integer count, which is set to 0 initially. Then, for each $t \in R_j$ in the last column, the count of array[t] is incremented by one. In other words, each $t \in R_j$ induces a hit at index *t* in the array, which increments the hit-count at array[t]. Henceforth, we call the array as 'hit []', and the algorithm as hit-index.

**Observation 3:** The method described above has two issues. First, as shown in lemma, each '*t*' in the set $R_j$ lie in the range $(1-M) \leq t \leq N-1)$. For example, $f_{-1}$ = 2 in the table suggest that a two character match may be found if P is aligned at location -1 in T. However, for integer t < 0 'hit [t]' does not exist. As a solution, we assume the initial index of the text file to be M rather than 0. This amplifies each *i* of the table 1 by M ensuring all $t > 0$ in the last column. Now, since each t is hyped by M, the hit-count at hit[t] represents the number of character matches at alignment location t-M rather than *t* in the text. That means the hit-count at hit[0] represents the number of characters that match when P is aligned at location –M in T. The second issue is the size of the array 'N', which is undesirable for the modern databases. In section 5 we have shown that how this issue can be resolved using an array of size 2M.

# 4 PATTERN PREPROCESSING

The algorithm given in section 5 requires all shifts of a character to be retrieved quickly from the pattern. To process the pattern, an array of pointers shift [max_ASCII+1] is used, where max_ASCII is the maximum possible ASCII code of a character in the alphabet, which is typically 127 or 255. Each pointer in the array points to a linked list that stores all shifts of a particular character in the pattern. Initially, all linked lists are empty. While reading the pattern from left to right, a node containing its shift is created. Then, based on the ASCII value of the character, the node is mapped to a particular linked list. For example, let 'm' be the shift of a character in the pattern that is being read, and let '*j*' be its ASCII value. For this character, a node containing integer 'm' is created and then inserted at the beginning of the linked list pointed by *shift[j]*.

**Example 4:** Let P=FCTHZCTZCF be a pattern of size 10. The ASCII codes of the pattern characters are:- F=70, C=67, T=84, H=72, and Z=90. The shift[] is shown in figure 1.

------------------------------------------------------------

Input: pattern characters. Output: A shift array, such that each cell of the array points to a list that stores all shifts of a particular character in the pattern.

------------------------------------------------------------

```
Let 'node' be a structure with two
fields: integer s, and node type
pointer *next
node *shift[max_ASCII +1]
integer j, m = 0
for  j = 0 to max_ASCII do
     shift[ j ] = NULL
end for
while ( Not end of the pattern )   do
       j = ASCII(patten character)
       node *ptr = new node
       ptr → s = m++
       ptr→ next = shift [j]
       shift[j] = ptr
end while
```

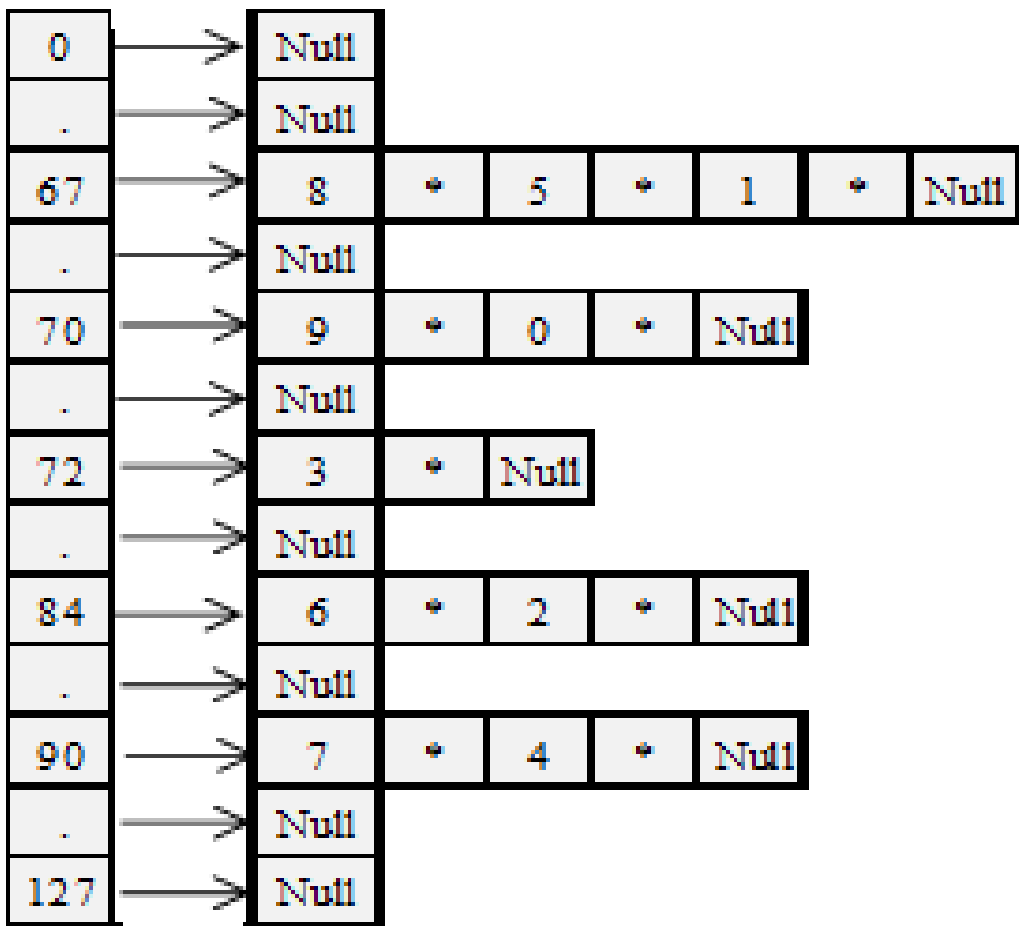

Figure 1: The Shift array.

# 5 ALGORITHM 'HIT-INDEX'

The challenge of the algorithm is to get the work done using an array of size 2M. As explained in the observation 3, the variable D is set to M, which is the presumed beginning of the text. With the arrival of each text character, D is incremented by 1. Corresponding to the size of the array 2M, the variable D is allowed to take values up to 3M-1, beyond which it is again reset to M. The algorithm reads a text character, and based on the ASCII value of the character, a particular list is chosen from the 'shift[]' array. For each shift 's' in the list, a hit-index '*t*' is computed, which is kept < 2M using modulus operator. Each '*t*' so computed increments the hit-count at 'hit[t]'.

---------------------------------------------------------------

Input: Pattern in the form of a 'shift[]' array, and a stream of text characters. Output: the number of characters that match when the pattern P is aligned from every location *i* in the text T such that $(1-M) \leq i \leq (N-1)$. The algorithm reads the text character by character, without any upper limit on N.

```
'node' is a structure with two fields:
integer s, and node type pointer *next
integer i=-M, D = M, j, t
integer hit[2M]

for j=0 to 2M-1  do
       hit[j]=0       /* initialize */
end for
/* read text*/
while(Not end of the text) do
       j = ASCII(text character)
       node *ptr = shift[j]
       while(ptr != NULL) do
              t = (D - (ptr → s))%2M
              hit[t]++
              ptr = ptr→ next
       end while
       Print i, hit[D-M]
       hit[D-M] = 0        /* reset */
       i++    D++
       if(D = = 3M)
              D = M
end while            /* text is over */
for j = 0 to M-1 do
       Print i, hit[D-M]
i++    D++
if(D = = 3M)
       D = M
end for
```

The printing of the hit-count in the array lags M locations behind the character being read. Therefore, the remaining M hit-counts are printed when the text input is over. Table 2 given below simulates the run time behavior of the algorithm for a pattern P=ABBA of size M=4. For the given pattern, the two linked lists in the shift array are: A= {3→0}, and B = {2→1}. The algorithm use an array integer of size 2M=8. All hit-counts in the array are initialized to 0. Let T be the text such that T = BBABAABBACAAB. The algorithm begins with printing the hit-counts from hit[0], which corresponds to the alignment location –M in the text. Since all t > 0, the hit-count at hit[0] is always 0, refer to observation 3. Therefore, for the first alignment at -M, the output is always 0. Each pair *(i, c)* in the last column represents the number of character matches '*c*' when P is aligned at location '*i*' in the text. Please view the table growing top-down with respect to each incoming text character. For each incoming text character, the algorithm performs three actions: a) generate the 'hit-indexes'(t), b) for each *t*, increment the hit-count at hit [t], and c) print, and then reset the hit-count at hit[D-M].

Table 2: Run Time Behavior.

| **i, T** | **D** | **t** | **Workspace: array hit [8]** | | | | | | | | **i, hit [D-M]** |
|---|---|---|---|---|---|---|---|---|---|---|---|
| | | | 0 | 1 | 2 | 3 | 4 | 5 | 6 | 7 | |
| | | | 0 | 0 | 0 | 0 | 0 | 0 | 0 | 0 | |
| -4 # | | | | | | | | | | | |
| -3 # | | | | | | | | | | | |
| -2 # | | | | | | | | | | | |
| -1 # | | | | | | | | | | | |
| 0 B | 4 | 2, 3 | ~~0~~ | 0 | 1 | 1 | 0 | 0 | 0 | 0 | -4, 0 |
| 1 B | 5 | 3, 4 | 0 | ~~0~~ | 1 | 2 | 1 | 0 | 0 | 0 | -3, 0 |
| 2 A | 6 | 3, 6 | 0 | 0 | ~~1~~ | 3 | 1 | 0 | 1 | 0 | -2, 1 |
| 3 B | 7 | 5, 6 | 0 | 0 | 0 | ~~3~~ | 1 | 1 | 2 | 0 | -1, 3 |
| 4 A | 8 | 5, 0 | 1 | 0 | 0 | 0 | ~~1~~ | 2 | 2 | 0 | 0, 1 |
| 5 A | 9 | 6, 1 | 1 | 1 | 0 | 0 | 0 | ~~2~~ | 3 | 0 | 1, 2 |
| 6 B | 10 | 0, 1 | 2 | 2 | 0 | 0 | 0 | 0 | ~~3~~ | 0 | 2, 3 |
| 7 B | 11 | 1, 2 | 2 | 3 | 1 | 0 | 0 | 0 | 0 | ~~0~~ | 3, 0 |
| 8 A | 4 | 1, 4 | ~~2~~ | 4 | 1 | 0 | 1 | 0 | 0 | 0 | 4, 2 |
| 9 C | 5 | --- | 0 | ~~4~~ | 1 | 0 | 1 | 0 | 0 | 0 | 5, 4 |
| 10A | 6 | 3, 6 | 0 | 0 | ~~1~~ | 1 | 1 | 0 | 1 | 0 | 6, 1 |
| 11A | 7 | 4, 7 | 0 | 0 | 0 | ~~1~~ | 2 | 0 | 1 | 1 | 7, 1 |
| 12B | 8 | 6, 7 | 0 | 0 | 0 | 0 | ~~2~~ | 0 | 2 | 2 | 8, 2 |
| # | 9 | | 0 | 0 | 0 | 0 | 0 | 0 | 2 | 2 | 9, 0 |
| # | 10 | | 0 | 0 | 0 | 0 | 0 | 0 | 2 | 2 | 10, 2 |
| # | 11 | | 0 | 0 | 0 | 0 | 0 | 0 | 2 | 2 | 11, 2 |
| # | 4 | | 0 | 0 | 0 | 0 | 0 | 0 | 2 | 2 | 12, 0 |

Let us consider first two characters of the text. For the first character ‘B’, the algorithm retrieves its shifts from the shift array to produce two hit-indexes t = 4 - 2 = 2, and t = 4–1 = 3. Hence, the hit-count at hit[2] and hit[3] is incremented from 0 to 1. Thereafter, the hit-count at hit[0] is printed, and then reset to 0. In the table, cells receiving hits are underlined, while the resetting is indicated by a strikethrough. For the next character ‘B’, the algorithm hits at locations 3 and 4. As a result, the hit-count at hit[3] and hit[4] is incremented to 2 and 1 respectively. For all other locations, the hit-count remains the same. This time, the algorithm prints the hit-count at hit [1], before it is reset to 0. As the process continues, D takes values up to 3M-1 (11), beyond which D is reset to M=4. The printout lags ‘M’ locations behind the character being read; therefore, the last M rows of the table show hit-counts without any input.

# 6 TIME AND SPACE ANALYSIS

In the preprocessing phase, insertion of a node in the beginning of the linked list costs *O(1)* time. Hence, for a pattern of size M, the pattern processing time is *O(M)*. Since a total of M nodes are inserted in the shift array of size max_ASCII + 1, the pattern preprocessing phase consumes *O(max_ASCII +M)* memory, where *max_ASCII* is the maximum ASCII code of a character in the alphabet, which is typically 128 or 256. An additional array of size 2M is used in the search phase, which gives the total run-time space requirement to be *O(max_ASCII + M) + O(2M)* i.e. *O(M)*. Thus, the total run-time memory requirement of the algorithm is independent of N, which is one of the desired issues when working with the large databases.

Let’s discuss the time consumed in the search phase. The outer while-loop of the algorithm hit-index given in section 5 reads a text character $T_i$, and then retrieves all shifts of $T_i$ from the shift array. That means for each text character $T_i$, the inner while-loop runs $fT_i$ times, where $fT_i$ is the frequency of $T_i$ in the pattern. Assuming a uniform character distribution, the average frequency of a character in the pattern can be given by $M / \sigma$, where, σ is the size of the alphabet. For a text of size N, this gives the average case execution time to be $O(N (M / \sigma))$. Therefore, for a pattern of size M such that $M \leq \sigma$, the expected execution time is *O(N)*. The worst case time *O(NM)* is reached in a rare situation when a single letter is repeated N and M times in both text and the pattern, e.g., T=AAAA, and P=AAA.

# 7 EXPERIMENTAL RESULTS

To conduct our experiments, we have used natural language as a dataset. For natural language, we use a plain text version of the eBook “Pride and Prejudice” (Austen, 1813) retrieved from the project Gutenberg. The file has 704146 characters (0.7 million approx.). Ten different patterns of varying sizes were chosen from the different locations in the file. Fig. 2(a) given below shows the total number of induced ‘hits’ for the corresponding size of the pattern. Each ‘hit’ can be treated as a character comparison. Notice, corresponding to the largest pattern size M=100, the number of induced ‘hits’ is roughly equal to 6N. The execution time to process these patterns is shown in fig. 2(b). The experiments were conducted on a 64 bit machine, Windows 8.1, Intel® Core™ i3-3120M, CPU @2.5 GHz, RAM 4 GB, using MinGW, GNU gcc version 4.6.2. The theoretical average execution time of the algorithm is $O(N (M / \sigma))$. Therefore, keeping the database unchanged, a 10 times increase in the size of the pattern, we expect a similar 10 times increase in the execution time. However, fig. 2(b) shows far more encouraging results. One final point, while conducting these experiments, the algorithm consumed just 0.3 MB of RAM space, which remains constant throughout the execution.

# 8 CONCLUSION

The completely new method we have proposed in this paper could be useful for researchers in the field of string matching. Text processing algorithms have seen widespread use of complex data structures. However, there are situations when the price paid in using these data structures dominates the overall gains. In this paper, we provide a highly practical algorithm that do not make use of such data structures, neither do we create indexes over the text, yet achieve *O(N)* average case time for patterns of size less than σ. Experiments have shown that larger patterns can also be dealt by the algorithm without performance deterioration. The proposed solutions are easy to implement with minimal effort and resources.

# ACKNOWLEDGEMENT

We thank the Ministry of Higher Education, Oman for providing with resources to conduct this work.

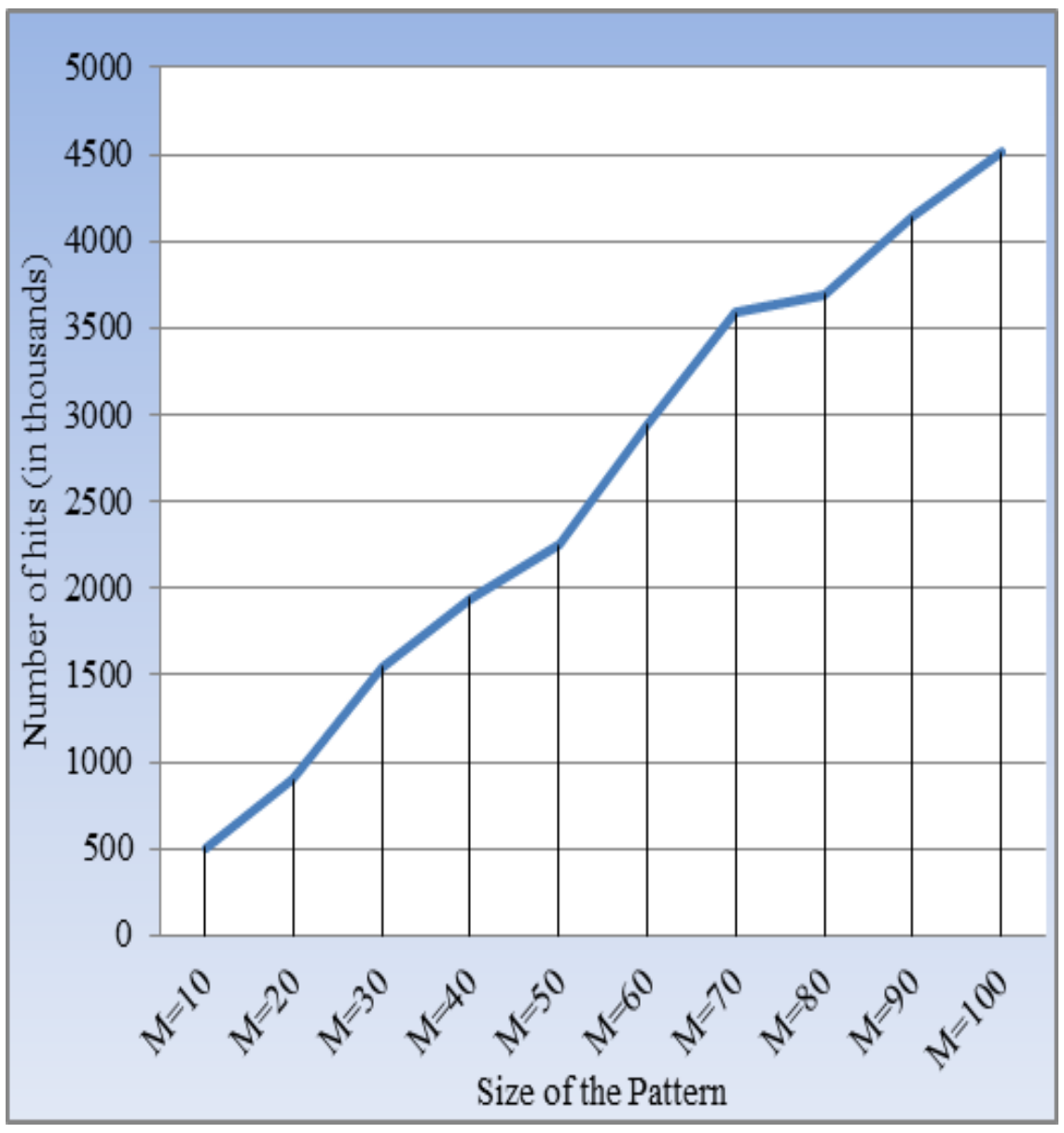

Figure 2(a): Number of hits (in thousands).

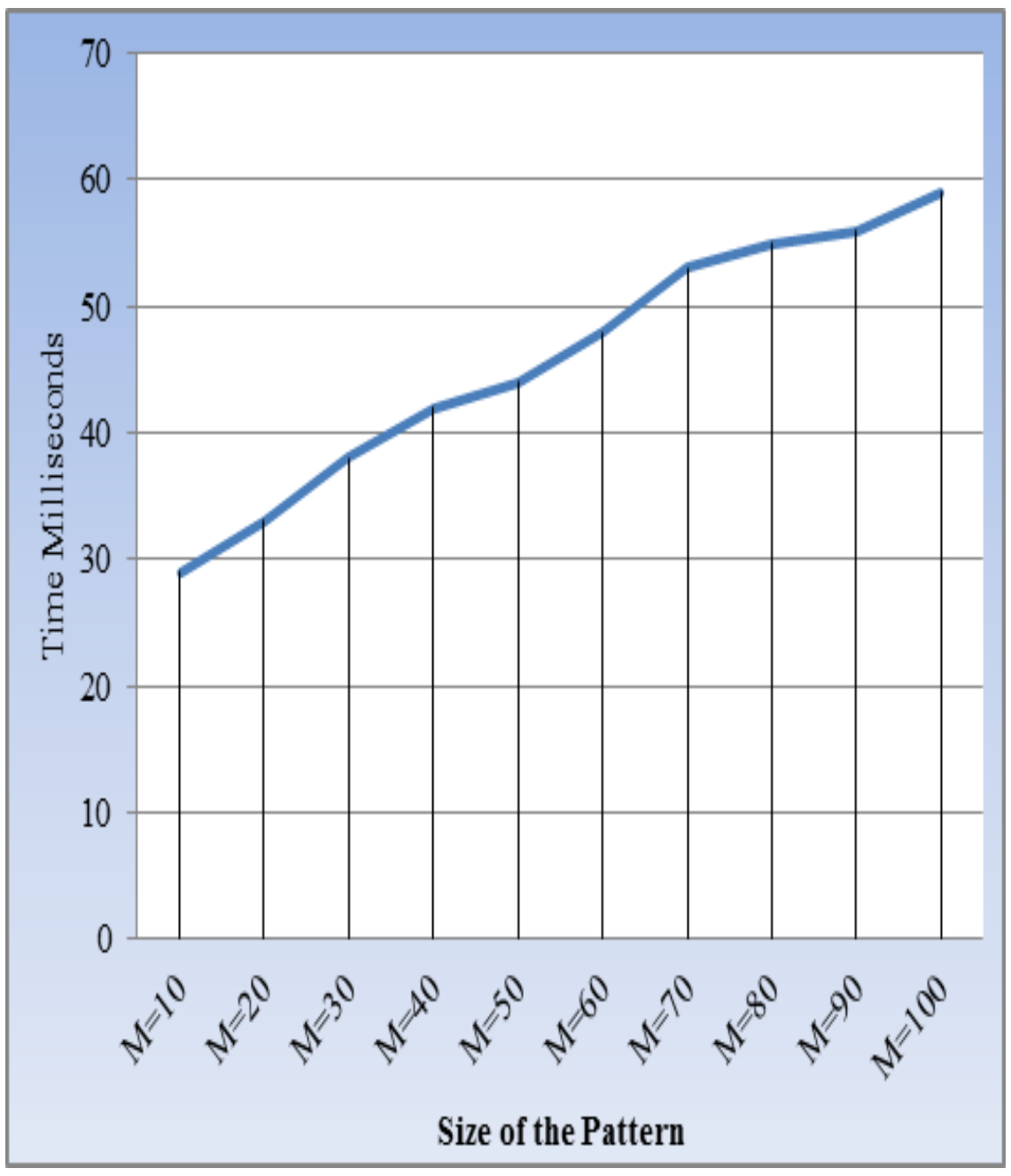

Figure 2(b): Execution Time.